\documentclass[preprint,12pt]{elsarticle}

\usepackage{amssymb}
 \biboptions{sort&compress}

\begin{document}

\pagestyle{empty}
\begin{frontmatter}



\title{Many to Many Matching with Demands and Capacities}


\author[la1]{Fatemeh Rajabi-Alni}

\corref{cor1}\ead {fatemehrajabialni@yahoo.com}
\cortext[cor1]{Corresponding author.}
\author[la2]{Alireza Bagheri}

 %\fntext[label2]{salam}

 \address[la1]{Department of Computer Engineering, Islamic Azad University,\\North Tehran Branch, Tehran, Iran.\fnref{label3}}

\address[la2]{Department of Computer Engineering and IT,\\Amirkabir University of Technology, Tehran, Iran.}

\begin{abstract}
Let $A$ and $B$ be two finite sets of points with total cardinality $n$, \textit {the many to many point matching with demands and capacities} matches each point $a_i \in A$ to at least $\alpha_i$ and at most ${\alpha '}_i$ points in $B$, and each point $b_j \in B$ to at least $\beta_j$ and at most ${\beta '}_j$ points in $A$ for all $1 \leq i \leq s$ and $1 \leq j \leq t$. In this paper, we present an upper bound for this problem using our new polynomial time algorithm.
\end{abstract}


\end{frontmatter}

% \linenumbers

\section{Introduction}
A \textit matching between two point sets $A$ and $B$ defines a relationship between them. The concept of the matching is used in various fields such as computational biology \cite{1}, pattern recognition \cite{2}, computer vision \cite{3}, music information retrieval \cite{4}, and computational music theory \cite{5}.
A \textit {many-to-many matching} between $A$ and $B$ maps each point of $A$ to at least one point of $B$ and vice-versa. Eiter and Mannila \cite{6} reduced the many-to-many matching problem to the minimum-weight perfect matching problem in a bipartite graph and solved it in $O(n^3)$ time. 

Consider two point sets $A=\{a_1,a_2,\dots,a_s\}$ and $B=\{b_1,b_2,\dots,b_t\}$ with $s+t=n$. Let $D_A=\{\alpha_1,\alpha_2,\dots,\alpha_s\}$ and $D_B=\{\beta_1,\beta_2,\dots,\beta_t\}$ denote the demand sets of $A$ and $B$, respectively. Let $C_A=\{{\alpha '}_1,{\alpha '}_2,\dots,{\alpha '}_s\}$ and $C_B=\{{\beta '}_1,{\beta '}_2,\dots,{\beta '}_t\}$ be the capacity sets of $A$ and $B$, respectively. \textit {The minimum-cost many-to-many matching with demands and capacities}, here called \textit {MMDC} matching, is a matching that matches each point $a_i\in A$ to at least $\alpha_i$ and at most ${\alpha '}_i$ points in $B$, and each point $b_j\in B$ to at least $\beta_j$ and at most ${\beta '}_j$ points in $A$, such that the sum of the matching costs is minimized. A function $\delta(a_i,b_j)$ represents the cost of the pairing $(a_i,b_j)$. Note that $\delta(a_i,b_j)$ can be zero, positive or negative. The cost of a matching is the sum of the costs of the pairings $(a_i,b_j )$ for all $1\leq i\leq s$ and $1\leq j\leq t$.

Schrijver \cite{7} proved that an MMDC matching can be found in strongly polynomial time. In this paper, we propose a new $O(n^6)$ algorithm for the MMDC matching problem based on the well known approach of Eiter and Mannila \cite{6} and provide an upper bound for it.

\section {MATCHING ALGORITHM}
\label{Matchingsec}

We construct a complete bipartite graph such that by applying the Hungarian method on it the demands and capacity limitations of the elements be satisfied. In the following, we explain how our complete bipartite graph $G$ is constructed.

We represent a set of the related nodes using a rectangle, each connection between two nodes with a line, and each node with a circle. So a connection between two nodes is shown using a line that connects the two corresponding circles. A directed line is used to show a branched line such that the input line branches into the output lines. \textit {A complete connection} between two sets is a connection where each node of one set is connected to the all nodes of the other set. We show a complete connection using a line connecting the two corresponding sets. 

Let $S \cup T$ be a bipartition of $G$, where $S=(\bigcup_{i=1}^{s} A_i) \cup (\bigcup_{i=1}^{s}{A'}_i ) \cup ( \bigcup_{j=1}^{t} X_j) \cup ( \bigcup_{j=1}^{t} W_j)$ and $T=\bigcup_{i=1}^{s} Bset_i$. The points of the sets $A_i$, $Bset_i$, and ${A'}_i$ for all $1\leq i\leq s$ are called the main points, since they are copies of the input points. On the other hand, the points of the sets $X_j$ and $W_j$ for all $1\leq j\leq t$ are called the dummy points. All edges $(a,b)$ that their both end points are main points, that is $a \in A_i \cup {A}'_i$ and $b \in Bset_i$ for $1\leq i\leq s$, are called the main edges.   

The Hungarian method computes a \textit {perfect} matching where each node is incident to a unique edge. We aim to find an MMDC matching that in which two or more nodes may be mapped to a same node, that is a node may be selected more than once. So our constructed graph contains multiple copies of each elements to simulate this situation. Let $A_i=\{a_{i1},\dots,a_{i\alpha_i }\}$ for $1\leq i\leq s$ be the set of the $\alpha_i$ copies of the point $a_i$. Each set $A_i$ is completely connected to the set $Bset_i= \{b_{1i},\dots,b_{ti}\}$ for $1\leq i\leq s$. This complete connection is shown using a line connecting the two corresponding rectangles of $A_i$ and $Bset_i$. We define a function $F_i$ for $A_i$ sets, such that $F_i(a_{ik},b_{ji})=\delta(a_i,b_j)$. Each $A_i$ set guarantees that each point $a_i \in A$ is matched to at least $\alpha_i$ elements of $B$.

Note that each node of $A$ has a limited capacity, i.e. it must be matched to at most a given number of the points of the other set. Each point $a_i$ is copied $({\alpha}'i-\alpha_i)$ times and constitute the ${A'}_i$ set. Let ${A'}_i=\{{a'}_{i1},\dots,{a'}_{i({\alpha}'i-\alpha_i)}\}$ for $1\leq i\leq s$. ${A'}_i$ sets guarantee that each point $b_j \in B$ is matched to at least $\beta_j$ elements of $A$. Moreover, each set ${A'}_i$ assures that each point $a_i$ is matched to at most ${\alpha}'_i$ elements. Each ${A'}_i$ is completely connected to $Bset_i$, where the cost of $({a'}_{id},b_{ji})$ edge is equal to $\delta(a_i,b_j)$ for all $1 \leq d \leq ({\alpha}'i-\alpha_i)$.

Assume that all nodes $b_{ji}$ for $1\leq i\leq s$ constitute sets, called $B_j$. In fact, the set $B_j$ is $s$ copies of $b_j$. We use the $W_j= \{w_{j1},\dots,w_{j(s-{\beta '}_j)}\}$ set to limit the number of the points that can be matched to $b_j \in B$ for $1 \leq j \leq t$. There is a zero wighted complete connection between the nodes of $B_j$ and $W_j$ for $1\leq j\leq t$.

Let $X_j=\{x_{j1},\dots,x_{j({\beta}'_j-\beta_j)}\}$ and $\gamma=\min\delta(a_i,b_j)\  for \  all\ 1\leq i\leq s\ and\  1 \leq j \leq t$. Select an arbitrary number ${\gamma}'$ such that ${\gamma}'< \gamma$, there exists a ${\gamma}'$ weighted complete connection between the nodes of $B_j$ and $X_j$ for all $1\leq j \leq t$. $X_j$ sets guarantee that the matching is a minimum cost matching.

There exists another set that compensates the bipartite graph, called $Y$. The input of the Hungarian algorithm is a complete bipartite graph, i.e. both parts of the input bipartite graph have an equal number of points. So we should balance the parts of our constructed bipartite graph before using the Hungarian algorithm. 
We have $$|S|=|\bigcup_{i=1}^{s} A_i| + |\bigcup_{i=1}^{s}{A'}_i|+|\bigcup_{j=1}^{t} X_j|+|\bigcup_{j=1}^{t} W_j|$$$$=\sum_{i=1}^s\alpha_i+\sum_{i=1}^s{\alpha}'_i-\sum_{i=1}^s\alpha_i+\sum_{j=1}^t{\beta '}_j-\sum_{j=1}^t\beta_j+s*t-\sum_{j=1}^t{\beta '}_j$$$$=\sum_{i=1}^s{\alpha}'_i+(s*t)-\sum_{j=1}^t\beta_j$$
, and $$|T|=|\bigcup_{i=1}^{s} Bset_i|=(s*t).$$ 

Let $|Y|=|\sum_ {i=1}^s\alpha '_i-\sum_{j=1}^t\beta_j|$, we add $Y$ to $S$ or $T$ depending on the values of $\sum_ {i=1}^s\alpha '_i$ and $\sum_{j=1}^t\beta_j$. 

Two cases arise: $\sum_ {i=1}^s\alpha '_i<\sum_{j=1}^t\beta_j$ and $\sum_ {i=1}^s\alpha '_i>\sum_{j=1}^t\beta_j$. In the first case $|S|<|T|$, so we add the $Y$ set to $S$.  There is a zero weighted complete connection between $Bset_i$ and $Y$ for all $1\leq i\leq s$.

Now assume that $\sum_ {i=1}^s\alpha '_i>\sum_{j=1}^t\beta_j$. In this case, the compensator set $Y$ is inserted to $T$. In this case, there is a complete connection between $X_j$ and $Y$ that in which the cost of the edges is an arbitrary number ${\gamma}''$ with $ {\gamma}' <{\gamma}''<\gamma $. Consequently, the first priority of the points of $X_j$ is the points of $B_j$ set. Moreover, ${A'}_i$ is completely connected to $Y$ with zero weighted edges. The complete bipartite graph that in which $\sum_ {i=1}^s\alpha '_i>\sum_{j=1}^t\beta_j$ is shown in Figure 1.

 \begin{figure}
\vspace{-5cm}
\hspace{-4cm}
\resizebox{1.4\textwidth}{!}{%
  \includegraphics{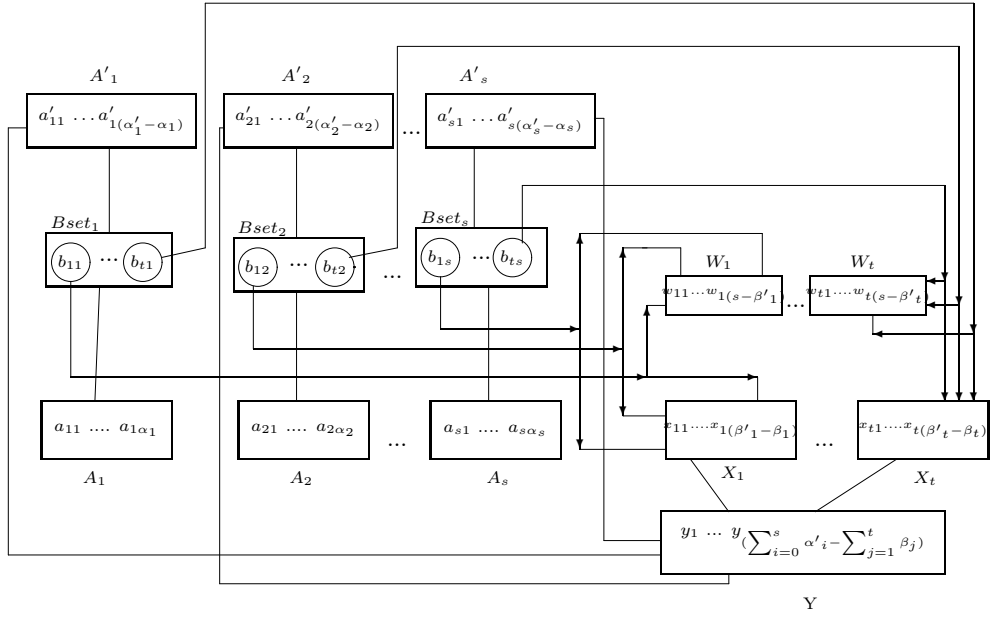}
}
%\vspace{5cm}       % Give the correct figure height in cm
\vspace{-15cm}
\caption{The constructed complete bipartite graph by our algorithm in which $\sum_ {i=1}^s\alpha '_i>\sum_{j=1}^t\beta_j$}
\label{fig:1}       % Give a unique label
\end{figure}

We claim that from a minimum weight perfect matching in $G=S \cup T$, we can get an MMDC matching between $A$ and $B$. Let $M$ be the union of the main edges of a minimum weight perfect matching in $G$. In the following, we prove that the weight of $M$ is equal to the cost of an MMDC matching between $A$ and $B$, called $L$.

%lemma 1
\newtheorem{lemma}{Lemma}
\begin{lemma}
$w(M)\leq c(L)$.
\end{lemma}
\textbf {Proof.} We get from $L$ a perfect matching $M'$ in our complete bipartite graph $G$, such that the cost of the union of the main edges $M''$ be equal the cost of $L$, that is $w(M'')=c(L)$. So we want to convert an MMDC matching between $A$ and $B$ to a perfect matching in $G$. In a perfect matching each vertex is incident to exactly one edge, so we relate an edge of $G$ to each pairing of $L$ as following.

Let $p_i$ be the number of the points $b_j \in B$ that are matched to $a_i\in A$ in the MMDC matching $L$. It is Obvious that $\alpha_i\leq p_i \leq {\alpha '}_i$. For each pairing $(a_i,b_j)$ in $L$, we connect $b_{ji}$ to one of unmatched points of $A_i$, that is $a_{ik}$ with $1\leq k \leq \alpha_i$. Then, depending on the value of $p_i$ two cases arise:
\begin{itemize}
\item
$p_i=\alpha_i$. In this situation, we add the zero weighted edges of $G$ that connect each ${a'}_{ij}\in {A'}_i$ to one of unmatched nodes of $Y$ for all $1\leq j \leq ({\alpha}'_i-\alpha_i)$.

\item $p_i>\alpha_i$. In this case, we need to match the $p_i-\alpha_i$ nodes of ${A'}_i$ with the nodes of $Bset_i$. So, for each pairing $(a_i,b_j)$ of the $p_i-\alpha_i$ remaining pairings we add an edge of $G$ that connects ${a'}_{ij}$ to $b_{ji}$. Then, if yet there exist nodes of ${A'}_i$ that have not been matched, for each of them we select a zero weighted edge of $G$ that connects it to an unmatched node of $Y$, and add it to the matching $M'$. 

\end{itemize}

Now, for each $w_{jk} \in W_j$ we add the edge of $G$ that connects it to an unmatched node of $B_j$. The points of $X_j$ are matched to the points of $B_j$, unless no points remain unmatched in $B_j$. So we first add the edges that connect the nodes of $X_j$ to the remaining unmatched nodes of $B_j$. Then, we add the edges that connect the unmatched nodes of $X_j$, if exists, to the unmatched nodes of $Y$.
 
Note that some points of $Bset_i$ sets for $1 \leq i \leq s$ may remain unmatched. We match the remaining unmatched points of the $Bset_i$ sets with the unmatched nodes of $Y$. Recall that this situation arises when $\sum_ {i=1}^s\alpha '_i<\sum_{j=1}^t\beta_j$. 

Since all nodes of $G$ are selected once, $M'$ is a perfect matching. For each $(a_i,b_j)\in L $ there is an edge with equal weight in $M''$, so $w(M'')=c(L)$. $M$ is the union of the main edges incident to the nodes of $M''$ in a minimum weight perfect matching in $G$, that is $w(M)\leq w(M'')$, so $w(M)\leq 
c(L)$.\qed

\begin{lemma}
 $w(M)\geq c(L)$.
\end{lemma}

\textbf {Proof.} From the union of the main edges $M$ of a minimum weight perfect matching in $G$, we get an MMDC matching $L'$ between $A$ and $B$, such that the cost of $L'$ be equal to the cost of $M$, that is $w(M)=c(L')$.
For each edge $m \in M$, if $m=(a_{ik},b_{ji} )$ or $m=({a'}_{ik},b_{ji} )$ then we add the pairing $(a_i,b_j )$ to $L'$. Otherwise, no pairing is added to $L'$. 

For each $a_i\in A$ for $1\leq i\leq s$, there exists the set $A_i$ in $G$ with $\alpha_i$ nodes which is connected only to one set, $Bset_i$. So the nodes of each $A_i$ are selected by some nodes of $Bset_i$, that is $b_{ji}$ for $1\leq j\leq t$. Hence, each $a_i\in A$ for $1\leq i\leq s$ is selected at least $\alpha_i$ times, so the demand of the point $a_i$ is satisfied. In $G$ there exist $\alpha_i$ plus ${\alpha}'_i-\alpha_i$ copies of each point $a_i$, that is the points of $A_i$ plus the points of $A'_i$. So the number of the points that are matched to each $a_i\in A$ is at most ${\alpha}'_i$.

Consider the sets $B_j$ with $1\leq j\leq t$, recall that $B_j=\{b_{ji}|1\leq i\leq s\}$ and the points of $W_j$ sets are connected to $B_j$ for $1\leq j\leq t$ by zero weighted edges. $W_j$ is connected only to $B_j$, so $W_j$ selects $s-{\beta '}_j$ number of the $s$ members of $B_j$ and ${\beta '}_j$ points remain unmatched in $B_j$. Suppose that $k$ points of ${\beta '}_j$ points in $B_j$ are selected by the points of $A_i$ sets for $1\leq i\leq s$, so the ${\beta '}_j-k$ remaining points of $B_j$ should be selected by the other sets that are connected to it. We discuss two cases, depending on the value of $k$.
\begin{itemize}
\item if $k<\beta_j$ then $({\beta '}_j-k)>({\beta '}_j-\beta_j)$. Then, $X_j$ selects the ${\beta '}_j-\beta_j$ elements of the remaining members of $B_j$. we have
$$({\beta '}_j-k)-({\beta '}_j-\beta_j)={\beta '}_j-k-{\beta '}_j+\beta_j=\beta_j-k>0,$$ so the remaining $\beta_j-k$ members of $B_j$ are selected by the points of ${A'}_i$ sets. Note that $k$ points of $b_{ji}$ points for all $1\leq i\leq s$ are selected by the points of $A_i$ sets and $\beta_j-k$ points of them are selected by ${A'}_i$ sets. The demand of the point $b_j$ is satisfied, since $\beta_j-k+k=\beta_j$.
\item if $k>\beta_j$ then $({\beta '}_j-k)<({\beta '}_j-\beta_j)$ and all the $({\beta '}_j-k)$ remaining members of $B_j$ are selected by the points of $X_j$, so the result is minimum cost result.
\end{itemize}

The cost of $L'$ is equal to the weight of $M$, since for each edge of $M$ we add a pairing with equal cost to $L'$, so $c(L')=c(M)$. On the other hand, $L'$ is a many to many matching that satisfies the demands and capacities of $A$ and $B$. $L$ is an MMDC matching between $A$ and $B$, so $c(L)\leq c(L')$. Thus $c(L)\leq c(M)$.\qed  

\newtheorem{theorem}{Theorem}
\begin{theorem}
Let $M$ be the union of the main edges of a minimum weight perfect matching in $G$, and let $L$ be an MMDC matching between $A$ and $B$. Then, $w(M)=c(L)$.
\end{theorem}
\textbf {Proof.} By Lemma 1 and Lemma 2 $w(M)\leq c(L)$ and $w(M)\geq c(L)$, respectively. So $w(M)=c(L)$.\qed

The time-complexity of the Hungarian method is $O(n^3 )$, where the number of the nodes of the input graph is $O(n)$ \cite{6}. The number of the nodes of our complete bipartite graph is $O(n^2)$, so the complexity of our algorithm is $O(n^6)$.
\section{Conclusion}
\label{ConclusionSect}
We presented an $O(n^6)$ algorithm for the many to many matching with demands and capacities. In this version of the many-to-many matching problem, the points of the two sets $A$ and $B$ are $n$-dimensional points, with $n\geq 1$. We can limit their dimensions to one or two dimensional. It is expected that the complexity of the $n$-dimensional matching problem will be reduced by exploiting the geometric information. So the one and two dimensional version of this problem remains open.

\end{document}